\documentclass[12pt]{article}
\usepackage{amsmath}
\usepackage{amssymb}
\usepackage{graphicx}
\usepackage{ifthen}
\usepackage{verbatim}

\newlength{\captionwidth}
\newsavebox{\tempbox}
\newcommand{\mycaption}[2]{%
\par\vspace{10pt}\sbox{\tempbox}{Figure #1: #2}%
\ifthenelse{\lengthtest{\wd\tempbox>\captionwidth}}%
{\sbox{\tempbox}{Figure.#1:\ }%
\addtolength{\captionwidth}{-\wd\tempbox}%
\mbox{Figure #1:\ }\parbox[t]{\captionwidth}{\small\textit{#2}}}%
{Figure #1: {\small\textit{#2}}}}%

\numberwithin{equation}{section}

\allowdisplaybreaks
\begin{document}
\thispagestyle{empty}
\begin{flushright}
June 2008\\
\end{flushright}
\bigskip
\bigskip

\begin{center}
{\large 
\textbf{Integrable Structure of 
        $5d$ $\mathcal{N}=1$ Supersymmetric Yang-Mills}}\\
{\large \textbf{and}}\\ 
{\large \textbf{Melting Crystal}
\footnote{\small 
Based on an invited talk presented at the international workshop 
{\it ``Progress of String Theory and Quantum Field Theory''} 
(Osaka City University, December 7-10, 2007).}} 
\end{center}
\bigskip
\bigskip

\renewcommand{\thefootnote}{\fnsymbol{footnote}}
\begin{center}
Toshio Nakatsu
\footnote{E-mail: \texttt{nakatsu@phys.sci.osaka-u.ac.jp}}$^1$, 
Yui Noma 
\footnote{E-mail: \texttt{yuhii@het.phys.sci.osaka-u.ac.jp}}$^1$
and 
Kanehisa Takasaki
\footnote{E-mail: \texttt{takasaki@math.h.kyoto-u.ac.jp}}$^2$\\
\bigskip
{\small
\textit{$^1$Department of Physics, Graduate School of Science,
Osaka University,\\
Toyonaka, Osaka 560-0043, Japan}}\\
{\small
\textit{$^2$Graduate School of Human and Environmental Studies, 
Kyoto University,\\ 
Yoshida, Sakyou, Kyoto 606-8501, Japan}}
\end{center}
\bigskip
\renewcommand{\thefootnote}{\arabic{footnote}}
\begin{abstract}
We study loop operators of $5d$ $\mathcal{N}=1$ SYM 
in $\Omega$ background. 
For the case of $U(1)$ theory, 
the generating function of correlation functions 
of the loop operators 
reproduces the partition function of 
melting crystal model with external potential.  
We argue the common integrable structure of 
$5d$ $\mathcal{N}=1$ SYM and melting crystal model.  
\end{abstract}

\setcounter{footnote}{0}
\newpage

\section{\large Introduction}

It is shown in \cite{Nekrasov-Okounkov} that
the Seiberg-Witten solutions \cite{Seiberg-Witten} 
of $4d$ $\mathcal{N}=2$ supersymmetric gauge theories  
emerge through \textit{random partition}, 
where 
Nekrasov's formulas \cite{Nekrasov-Okounkov,Nekrasov} 
for these gauge theories are 
understood as the partition functions of random partition. 
The integrable structure of random partition 
is elucidated in \cite{Marshakov-Nekrasov}, and 
thereby the integrability of correlation functions 
among single-traced chiral observables is explained. 
Such an extension of the Seiberg-Witten geometries 
also becomes attractive to understand $4d$ $\mathcal{N}=1$ 
supersymmetric gauge theories by providing a powerful tool 
\cite{Itoyama}.

Integrable structure of melting crystal model 
with external potential is clarified  
in \cite{Nakatsu-Takasaki}. 
Melting crystal model, 
known as 
\textit{random plane partition} 
has a significant relation with 
$5d$ $\mathcal{N}=1$ supersymmetric gauge theories. 
Nekrasov's formula 
for these gauge theories  
can be retrieved from the partition function 
of melting crystal model \cite{MNTT1}, 
where the model is interpreted as a $q$-deformed random partition. 
It is argued \cite{Nakatsu-Takasaki} 
a relation between 
loop operators of $5d$ $\mathcal{N}=1$ 
supersymmetric Yang-Mills (SYM) 
and external potentials of the melting crystal model.

We start Section 2 with 
providing a brief review about $5d$ $\mathcal{N}=1$ SYM 
in $\Omega$ background \cite{Losev-Marshakov-Nekrasov}. 
We introduce loop operators of this theory. 
Computation of correlation functions 
among these operators  is discussed. 
Generating function of the correlation functions of $U(1)$ theory
reproduces the partition function of the aforementioned 
melting crystal model. 
In Section 3 we discuss a common integrable structure 
of $5d$ $\mathcal{N}=1$ SYM in $\Omega$ background 
and melting crystal model for the case of the $U(1)$ theory. 
In Section 4 we present an extension of 
the Seiberg-Witten geometry of the $U(1)$ theory 
by using the loop operators.

\section{\large 
Loop operators of $5d$ $\mathcal{N}=1$ SYM in $\Omega$ background}

We first consider 
an ordinary $5d$ $\mathcal{N}=1$ SYM on $\mathbb{R}^4\times S^1$. 
Let $E$ be the $SU(N)$-bundle on $\mathbb{R}^4$ with $c_2(E)=n \geq 0$. 
A gauge bundle of this theory is the $SU(N)$-bundle $\pi^*E$ on 
$\mathbb{R}^4\times S^1$ pulled back from $\mathbb{R}^4$. 
$\pi$ is the projection from 
$\mathbb{R}^4\times S^1$ to $\mathbb{R}^4$.  
All the fields in the vector multiplet 
are set to be periodic along $S^1$. 
The bosonic ingredients are a $5d$ gauge potential $A_M(x,t)dx^M$ 
and a scalar field $\varphi(x,t)$ 
taking the value in $su(N)$. 
These describe a $5d$ Yang-Mills-Higgs system.  
The gauge potential can be separated into two parts 
$A_{\mu}(x,t)dx^{\mu}$ and $A_t(x,t)dt$, 
respectively the components 
of the $\mathbb{R}^4$- and the $S^1$-directions. 
Let $\mathcal{A}_E$ be the infinite dimensional affine space 
consisting of all the gauge potentials on $E$. 
$A_{\mu}(x,t)dx^{\mu}$ describes a loop $A(t)$ in $\mathcal{A}_E$, 
where the loop is parametrized by the fifth-dimensional circle. 
As for $A_t(x,t)$, together with $\varphi(x,t)$, 
the combination $A_t+i\varphi$ describes a loop $\phi(t)$ 
in $\Omega^0(\mathbb{R}^4,\mbox{ad}E \otimes \mathbb{C})$, 
the space of all the sections of 
$\mbox{ad}E \otimes \mathbb{C}$, 
where ad$E$ is the adjoint bundle on $\mathbb{R}^4$ with fibre $su(N)$. 
Taking account of the periodicity,  
the same argument is also applicable to the gauginos. 
The vector multiplet thereby describes 
a loop in the configuration space of the $4d$ theory. 
In the case of the Yang-Mills-Higgs system, 
the loop $A(t)$ gives 
a family of covariant differentials on $E$ as  $d_{A(t)}=d+A(t)$.   
For the loop $\phi(t)$, 
since it involves $A_t(x,t)$, 
it becomes convenient to introduce the differential operator 
\begin{eqnarray}
\mathcal{H}(t)\equiv 
\frac{d}{dt}+\phi(t)\,. 
\label{H(t)}
\end{eqnarray}

\subsection{ \normalsize 
$5d$ $\mathcal{N}=1$ SYM in $\Omega$ background}

Via the standard dimensional reductions,
$6d$ $\mathcal{N}=1$ SYM gives 
lower dimensional Yang-Mills theories with $8$ supercharges, 
including the above theory. 
Furthermore, 
the dimensional reductions 
in the $\Omega$ background 
provide powerful tools 
to understand these theories 
\cite{Losev-Marshakov-Nekrasov}.
The $\Omega$ background is 
a $6d$ gravitational background 
on $\mathbb{R}^4 \times T^2$ 
described by a metric of the form: 
$
ds^2=
\sum_{\mu=1}^4
(dx^{\mu}-\sum_{a=5,6}V_a^{\mu}dx^a)^2+\sum_{a=5,6}(dx^a)^2\,, 
$
where two vectors $V_5^{\mu},V_6^{\mu}$ 
generate rotations on two-planes $(x^1,x^2)$ and $(x^3,x^4)$ 
in $\mathbb{R}^4$.  
By letting  
$V_1=x^2\frac{\partial}{\partial x^1}-x^1\frac{\partial}{\partial x^2}$ 
and 
$V_2=x^4\frac{\partial}{\partial x^3}-x^3\frac{\partial}{\partial x^4}$, 
they are respectively the real part and the imaginary part of 
the combination    
\begin{eqnarray}
V_{\epsilon_1,\epsilon_2}
&\equiv& \epsilon_1V_1+\epsilon_2V_2\,,  
\hspace{6mm}\epsilon_1,\epsilon_2 \in \mathbb{C}.
\label{V_epsilon}
\end{eqnarray}
The above combination is expressed in component as 
$V_{\epsilon_1,\epsilon_2}
=\Omega^\mu_{~\nu} x^\nu\frac{\partial}{\partial x^\mu}$.

To see the dimensional reduction in the $\Omega$-background, 
we first consider the bosonic part of the $5d$ SYM. 
The corresponding Yang-Mills-Higgs system is modified 
from the previous one.  
However, 
the system is eventually controlled by 
replacing $\mathcal{H}(t)$ with  
\begin{eqnarray}
\mathcal{H}_{\epsilon_1,\epsilon_2}(t)
\equiv 
\mathcal{H}(t)+\mathcal{K}_{\epsilon_1,\epsilon_2}(t)\,.
\label{H_epsilon}
\end{eqnarray}
Here $\mathcal{K}_{\epsilon_1,\epsilon_2}(t)$ 
is an another differential operator of the form 
\cite{student}
\begin{eqnarray}
\mathcal{K}_{\epsilon_1,\epsilon_2}(t)
\equiv 
V_{\epsilon_1,\epsilon_2}^\mu \partial_{A(t)\,\mu}
+
\frac{1}{2}\Omega^{\mu \nu}\mathcal{J}_{\mu \nu}\,, 
\label{K_epsilon}
\end{eqnarray}
where 
$\mathcal{J}_{\mu \nu}$ denote 
the $SO(4)$ Lorentz generators of the system. 
This operator generates a $T^2$-action  
by taking the commutators with $d_{A(t)}$ and $\mathcal{H}(t)$. 
For instance, we have 
\begin{eqnarray} 
[d_{A(t)}, \mathcal{K}_{\epsilon_1,\epsilon_2}(t)]
=-\iota_{V_{\epsilon_1,\epsilon_2}}F_{A(t)}\,.
\label{torus action on A}
\end{eqnarray} 
The right hand side is precisely 
the transformation brought about on $\mathcal{A}_E$ 
by the infinitesimal rotation 
$\delta x^\mu=-V^{\mu}_{\epsilon_1,\epsilon_2}$.

The supercharges $Q_{\alpha a}$ and $\bar{Q}^{\dot{\alpha}}_{a}$ 
are realized in a way different from the case 
of $\epsilon_1=\epsilon_2=0$.  
Note that we use the $4d$ notation 
such that $\alpha,\dot{\alpha}$ and $a$ 
denote the indices of the Lorentz group $SU(2)_L\times SU(2)_R$ 
and the R-symmetry $SU(2)_I$. 
By the standard argument, 
we may interpret the $5d$ SYM as a topological field theory. 
Actually, 
by regarding the diagonal $SU(2)$ of $SU(2)_R\times SU(2)_I$ as 
a new $SU(2)_R$, we can extract a supercharge that behaves as 
a scalar under the new Lorentz symmetry. 
We write the scalar supercharge as $Q_{\epsilon_1,\epsilon_2}$. 
The gaugino acquires a natural interpretation  
as differential forms,  
$\eta(x,t),\psi_{\mu}(x,t)$ and $\xi_{\mu \nu}(x,t)$. 
These give 
fermionic loops, $\eta(t), \psi(t)$ and $\xi(t)$.  
The main part of the $Q$-transformation takes the forms
\begin{eqnarray}
&&
Q_{\epsilon_1,\epsilon_2}A(t)=\psi(t)\,, 
\hspace{7mm}
Q_{\epsilon_1,\epsilon_2}\psi(t)=[d_{A(t)}, 
\mathcal{H}_{\epsilon_1,\epsilon_2}(t)]\,,
\label{Q transform (A, psi)}
\\
&&
Q_{\epsilon_1,\epsilon_2}
\mathcal{H}_{\epsilon_1,\epsilon_2}(t)=0\,,
\label{Q transform H}
\end{eqnarray}
where $\psi(t)$ is a fermionic loop in 
$\Omega^1(\mathbb{R}^4,\mbox{ad}E)$.

\subsection{\normalsize 
Loop operators and their correlation functions}

Taking account of the relation $\phi(x,t)=A_t(x,t)+i\varphi(x,t)$, 
the following path-ordered integral 
provides an analogue of a holonomy of the gauge potential.
\begin{eqnarray}
W^{(0)}(x;t_1,t_2)=\mbox{P}e^{-\int_{t_2}^{t_1}dt\phi(x,t)}\,, 
\label{W_(0)}
\end{eqnarray}
where the symbol means the path-ordered integration, 
more precisely, 
it is defined by the differential equation  
\begin{eqnarray}
(\frac{d}{dt_1}+\phi(x,t_1))W^{(0)}(x;t_1,t_2)=0\,, 
\hspace{5mm}
W^{(0)}(x;t_2,t_2)=1\,. 
\label{def O_(0)}
\end{eqnarray} 
The trace of the holonomy along the circle 
defines a loop operator as 
\begin{eqnarray}
\mathcal{O}^{(0)}(x)= 
\mbox{Tr}\, W^{(0)}(x; R,0)\,, 
\label{O_(0)}
\end{eqnarray}
where $R$ is the circumference of $S^1$. 
The above operator is an analogue of 
the Wilson loop along the circle. 
Unlike the case of $\epsilon_1=\epsilon_2=0$, 
it is not $Q$-closed except at $x=0$. 
To see this, note that the $Q$-transformations 
(\ref{Q transform (A, psi)}) and (\ref{Q transform H}) 
imply    
$Q_{\epsilon_1,\epsilon_2}\phi(t)=
-\iota_{V_{\epsilon_1,\epsilon_2}}\psi(t)$. 
By using this, we find 
\begin{eqnarray}
Q_{\epsilon_1,\epsilon_2}\mathcal{O}^{(0)}(x)
=
\int_0^{R}dt_1 
\mbox{Tr}\Bigl\{\,W^{(0)}(x;\,R,t_1)\,
\iota_{V_{\epsilon_1,\epsilon_2}}\psi(x,t_1)\,
W^{(0)}(x;\,t_1,0)\Bigr\}\,. 
\label{QO_(0)}
\end{eqnarray}
Since the right hand side of the above formula vanishes only at $x=0$,  
this means that $\mathcal{O}^{(0)}(x)$ becomes $Q$-closed only at $x=0$.

The above property may be explained 
in terms of the equivariant de Rham theory. 
To see this, 
let us first generalize the path-ordered integral (\ref{W_(0)}) 
by exponentiating the combination 
$F_{A(t)}-\psi(t)+\phi(t)$ in place of $\phi(t)$ as 
\begin{eqnarray}
W(x;\,t_1,t_2)
=\mbox{P}e^{-\int_{t_2}^{t_1}dt \big(F_{A(t)}-\psi(t)+\phi(t)\big)(x)}\,, 
\label{W}
\end{eqnarray}
where the right hand side is given by a differential equation 
similar to (\ref{def O_(0)}). 
This means that $W$ has the components, 
according to degrees of differential forms on $\mathbb{R}^4$, 
as $W=W^{(0)}+W^{(1)}+\cdots+W^{(4)}$, 
where the indices denote the degrees.
We generalize the loop operator (\ref{O_(0)}) as 
\begin{eqnarray}
\mathcal{O}(x)=
\mbox{Tr}\, W(x; R,0)\,.
\label{O}
\end{eqnarray}
This also has components as $\mathcal{O}=\mathcal{O}^{(0)}
+\mathcal{O}^{(1)}+\cdots+\mathcal{O}^{(4)}$.  
Eq. (\ref{QO_(0)}) can be now expressed as 
$Q_{\epsilon_1,\epsilon_2}\mathcal{O}^{(0)}
=\iota_{V_{\epsilon_1,\epsilon_2}}\mathcal{O}^{(1)}$. 
This is actually the first equation 
among a series of the equations that $\mathcal{O}^{(i)}$ obey. 
Such equations eventually show up 
by expanding the identity \cite{Nakatsu-Noma-Takasaki} 
\begin{eqnarray}
(d_{\epsilon_1,\epsilon_2}+Q_{\epsilon_1,\epsilon_2})\mathcal{O}(x)=0\,, 
\label{formula of O}
\end{eqnarray}
where $d_{\epsilon_1,\epsilon_2}\equiv 
d-\iota_{V_{\epsilon_1,\epsilon_2}}$
is the $T^2$-equivariant differential on $\mathbb{R}^4$.

We can also consider the loop operators 
encircling the circle many times. 
Correspondingly we introduce 
\begin{eqnarray}
\mathcal{O}_k(x)=
\mbox{Tr}\,W(x;kR,0)\,, 
\hspace{6mm}
k=1,2,\cdots 
\label{O_k} 
\end{eqnarray}
These satisfy 
\begin{eqnarray}
(d_{\epsilon_1,\epsilon_2}+
Q_{\epsilon_1,\epsilon_2})\mathcal{O}_k(x)=0\,. 
\label{formula of O_k}
\end{eqnarray}

Let us examine the correlation functions 
$\langle\, 
\prod_{a}\int_{\mathbb{R}^4}\mathcal{O}_{k_a}\,
\rangle^{\epsilon_1,\epsilon_2}$.
Since the integral
$\int_{\mathbb{R}^4}\mathcal{O}_k=\int_{\mathbb{R}^4}\mathcal{O}_k^{(4)}$
is $Q$-closed by virtue of the formula (\ref{formula of O_k}),
these can be computed by a supersymmetric quantum mechanics (SQM) 
which is substantially equivalent to 
the $5d$ SYM as the topological field theory.
Such a SQM turns to 
be $T^2$-equivariant SQM on $\tilde{\mathcal{M}}_n$
\cite{Nekrasov},
where $\tilde{\mathcal{M}}_n$ 
is the moduli space of the framed $n$ instantons.
The $Q$-transformation 
(\ref{Q transform (A, psi)}) 
is converted to the supersymmetry of the quantum mechanics  
\begin{eqnarray}
Q_{\epsilon_1,\epsilon_2}m(t)=\chi(t)\,, 
\hspace{7mm}
Q_{\epsilon_1,\epsilon_2}\chi(t)=
-\frac{dm(t)}{dt}+\mathcal{V}_{\epsilon_1,\epsilon_2}(m(t))\,, 
\label{Q transform (m,chi)}
\end{eqnarray}
where $\mathcal{V}_{\epsilon_1,\epsilon_2}$ is 
the Killing vector 
induced by the variation 
$\delta A=\iota_{V_{\epsilon_1,\epsilon_2}}F_A$ on $\mathcal{A}_E$.
The combination $F_{A(t)}-\psi(t)+\phi(t)$ can be 
identified with a loop space analogue of 
the $T^2$-equivariant curvature $\mathcal{F}_{\epsilon_1,\epsilon_2}$ 
of the universal connection, 
where the universal bundle becomes equivariant by 
the $T^2$-action on $\mathcal{A}_E \times \mathbb{R}^4$.

In the computation of the correlation function, 
by virtue of the supersymmetry (\ref{Q transform (m,chi)}),  
only the constant modes $m_0,\chi_0$ contribute to 
the observable, 
and the above combination precisely becomes 
$\mathcal{F}_{\epsilon_1,\epsilon_2}$
\cite{Losev-Marshakov-Nekrasov}.
This means that $\mathcal{O}_k(x)$ truncates 
to the equivariant Chern character 
$\mbox{Tr} \, e^{-kR\, \mathcal{F}_{\epsilon_1,\epsilon_2}}$.
Thus we obtain the finite dimensional integral representation 
\begin{eqnarray}
\Big\langle 
\,\prod_{a}\int_{\mathbb{R}^4}\mathcal{O}_{k_a}\,
\Big \rangle_{n-instanton}^{\epsilon_1,\epsilon_2}
=
\frac{1}{(2\pi i R)^{ \frac{\dim \tilde{\mathcal{M}}_n}{2} }}
\int_{\tilde{\mathcal{M}}_n}
\hat{A}_{T^2}(R\,{\bf t}_{\epsilon_1,\epsilon_2},
\,\tilde{\mathcal{M}}_n)\,
\prod_{a}
\int_{\mathbb{R}^4} 
\mbox{Tr} \, 
e^{-k_aR \, \mathcal{F}_{\epsilon_1,\epsilon_2}}\,. 
\label{correlator of O_a}
\end{eqnarray}
where $\hat{A}_{T^2}(\cdot\,,\tilde{\mathcal{M}}_n)$ is 
the $T^2$-equivariant $\hat{A}$-genus of the tangent bundle 
of $\tilde{\mathcal{M}}_n$, 
and 
$\bf{t}_{\epsilon_1,\epsilon_2}$ 
is a generator of $T^2$ that gives the Killing vector 
$\mathcal{V}_{\epsilon_1,\epsilon_2}$.

Introducing the coupling constants $t=(t_1,t_2,\cdots)$,  
the generating function of the correlation functions 
is given by 
$\mathcal{Z}_{\epsilon_1,\epsilon_2}(t)
=\left \langle 
e^{\sum_{k}t_k \int_{\mathbb{R}^4}\mathcal{O}_k} 
\right \rangle^{\epsilon_1,\epsilon_2}$. 
Since $n$-instanton 
contributes with the weight $(R\Lambda)^{2nN}$, 
where $\Lambda$ is the dynamical scale,  
letting $Q=(R\Lambda)^2$, 
we can express the generating function as
\begin{eqnarray}
\mathcal{Z}_{\epsilon_1,\epsilon_2}(t)
=
\sum_{n=0}
Q^{nN}
\left \langle 
e^{\sum_{k}t_k \int_{\mathbb{R}^4}\mathcal{O}_k} 
\right \rangle_{n-instanton}^{\epsilon_1,\epsilon_2}  \,. 
\label{generating function SU(N)}
\end{eqnarray}

\subsection{\normalsize 
Application of localization technique}

The right hand side of 
the formula (\ref{correlator of O_a})
is eventually replaced with a statistical sum over partitions. 
To see their appearance, 
note that the integration 
localizes to the fixed points of the $T^2$-action. 
However, 
the fixed points in $\tilde{\mathcal{M}}_n$ 
are small instanton singularities  
since the variation 
$\delta A= -\iota_{V_{\epsilon_1,\epsilon_2}}F_A$ 
vanishes there. 
These can be resolved by instantons 
on a non-commutative $\mathbb{R}^4$. 
Applying such a regularization via the non-commutativity, 
the fixed points get isolated, 
so that they are eventually labelled by using partitions 
\cite{Nakajima}.

A partition $\lambda=(\lambda_1,\lambda_2,\cdots)$ is 
a sequence of nonnegative integers 
satisfying $\lambda_i \geq \lambda_{i+1}$ 
for all $i \geq 1$. 
Partitions are identified with the Young diagrams 
in the standard manner. 
The size is defined by $|\lambda|=\sum_{i \geq 1}\lambda_i$, 
which is the total number of boxes of the diagram.

Let us describe the formula (\ref{correlator of O_a}) 
for the $U(1)$ theory. 
The relevant computation of the localization 
can be found in \cite{Nakajima,Nakajima-Yoshioka-lec}. 
We truncate $\epsilon_{1,2}$ as 
$-\epsilon_{1}=\epsilon_2=i\hbar$, 
where $\hbar$ is a positive real parameter. 
Consequently, 
the formula becomes a $q$-series, 
where $q=e^{-R\hbar}$.  
The fixed points in $\tilde{\mathcal{M}}_n$ 
are labelled by partitions of $n$. 
The equivariant $\hat{A}$-genus takes the following form 
at the partition $\lambda$ of $n$: 
\begin{eqnarray}
(2\pi iR)^{-2n}
\left. 
\hat{A}_{T^2}(R\,{\bf t}_{-i\hbar,i\hbar}
\,\tilde{\mathcal{M}}_n)
\right |_{\lambda}
=
(-)^{n}
\Bigl(\frac{\hbar}{2\pi}\Bigr)^{2n}
\Bigl(\prod_{s \in \lambda}h(s)\Bigr)^2
q^{\frac{\kappa(\lambda)}{2}}
s_{\lambda}(q^{\rho})^2\,, 
\label{U(1) Dirac_index}
\end{eqnarray}
where $h(s)$ denotes the hook length of the box $s$ of 
the Young diagram $\lambda$, 
and 
$s_{\lambda}(q^{\rho})$ is the Schur function 
$s_{\lambda}(x_1,x_2,\cdots)$ specialized to $x_i=q^{i-\frac{1}{2}}$.
Similarly,  
the fixed points in $\tilde{\mathcal{M}}_n \times \mathbb{R}^4$ 
are labelled by partitions of $n$. 
Denoting them as $(\lambda,0)$,  
the equivariant Chern character takes the form  
$
\left.
\mbox{Tr}\,e^{-kR \mathcal{F}_{-i\hbar,i\hbar}}
\right|_{(\lambda,0)}
=\mathcal{O}_k(\lambda)
$, 
where $\mathcal{O}_k(\lambda)$ is given by 
\begin{eqnarray}
\mathcal{O}_{k}(\lambda)
=
(1-q^{-k})
\sum_{i=1}^{\infty}
\Bigl\{
q^{k(\lambda_i-i+1)}-q^{k(-i+1)}
\Bigr\}
+1\,. 
\label{O_k(lambda)}
\end{eqnarray}
The above functions have been exploited in 
\cite{Marshakov-Nekrasov, Kanno-Moriyama} 
from the $4d$ gauge theory viewpoint. 
By taking account of 
(\ref{U(1) Dirac_index}) and (\ref{O_k(lambda)}), 
the formula (\ref{correlator of O_a}) 
becomes eventually as 
\begin{eqnarray}
\Big\langle 
\,\prod_{a}\int_{\mathbb{R}^4}\mathcal{O}_{k_a}\,
\Big \rangle_{n-instanton}^{-i\hbar,i\hbar}
=
(-)^n
\sum_{|\lambda|=n}
q^{\frac{\kappa(\lambda)}{2}}
s_{\lambda}(q^{\rho})^2
\prod_{a}
\hbar^{-2}
\mathcal{O}_{k_a}(\lambda)\,.
\label{correlator of O_a_U(1)}
\end{eqnarray}

Although we have not taken into account, 
the Chern-Simon term can be added 
to a $5d$ gauge theory,   
with the coupling constant being quantized, 
in particular,  
for the $U(1)$ theory, 
$m=0,\pm 1$. 
It modifies  
the right hand side of (\ref{correlator of O_a_U(1)}) 
by giving a contribution of the form 
$(-)^{m|\lambda|}q^{-\frac{m\kappa(\lambda)}{2}}$,  
for each $\lambda$ \cite{MNTT1}.  
Hereafter, 
we consider the case of 
the $U(1)$ theory having  
the Chern-Simon coupling, $m=1$. 
The corresponding generating function becomes  
\begin{eqnarray}
\mathcal{Z}^{U(1)}_{-i\hbar,i\hbar}(t)
=
\sum_{\lambda} 
Q^{|\lambda|} 
s_{\lambda}(q^{\rho})^2
e^{\hbar^{-2}\sum_{k=1}t_k \mathcal{O}_k(\lambda)}\,.
\label{Z_U(1)}
\end{eqnarray}

\section{\large 
Integrability of  $5d$ $\mathcal{N}=1$ SYM in $\Omega$ background}

We can view the generating function (\ref{Z_U(1)}) 
as a $q$-deformed random partition. 
To see this,  
note that the $4d$ limit $R \rightarrow 0$ makes 
$q=e^{-R\hbar} \rightarrow 1$, 
the Boltzmann weight takes at this limit,  
the form    
$(\Lambda/\hbar)^{2|\lambda|}
\bigl(\prod_{s \in \lambda}h(s)\bigr)^{-2}$, 
which is the standard weight of a random partition. 
It can be also viewed as a melting crystal model, 
known as random plane partition.  
The corresponding model 
is studied in \cite{Nakatsu-Takasaki} 
as a melting crystal model with external potential, 
where the Chern characters $\mathcal{O}_k$ correspond 
precisely to the external potentials.

\subsection{\normalsize 
Melting crystal model}

A plane partition $\pi$ is an array of 
non-negative integers 
\begin{eqnarray}
\begin{array}{cccc}
\pi_{11} & \pi_{12} & \pi_{13} & \cdots \\
\pi_{21} & \pi_{22} & \pi_{23} & \cdots \\
\pi_{31} & \pi_{32} & \pi_{33} & \cdots \\
\vdots & \vdots & \vdots & ~
\end{array}
\label{pi}
\end{eqnarray}
satisfying 
$\pi_{ij}\geq \pi_{i+1 j}$ and $\pi_{ij}\geq \pi_{i j+1}$ 
for all $i,j \geq 1$. 
Plane partitions are identified 
with the $3d$ Young diagrams. 
The $3d$ diagram $\pi$ 
is a set of unit cubes such that $\pi_{ij}$ cubes 
are stacked vertically on each $(i,j)$-element of $\pi$. 
%
%
\begin{figure}[htb]
\begin{center}
\includegraphics[scale=0.43]{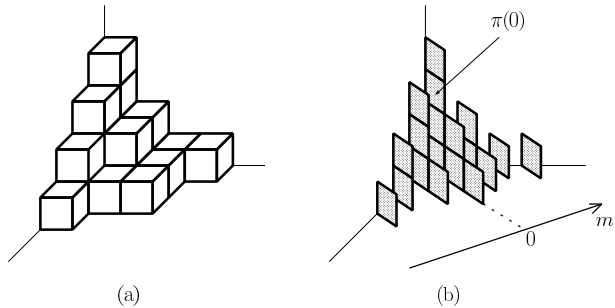}
\caption{\textit{The $3d$ Young diagram (a) 
and the corresponding sequence of partitions 
 (b).}}
\end{center}
\label{three-dimensional Young diagram}
\end{figure}
%
%
Diagonal slices of $\pi$ become partitions, 
as depicted in Fig.1.  
Denote $\pi(m)$ the partition along the $m$-th diagonal slice, 
where $m \in \mathbb{Z}$. 
In particular, 
$\pi(0)=(\pi_{11},\pi_{22},\cdots)$ 
is the main diagonal one.  
This series of partitions satisfies the condition
\begin{eqnarray}
\cdots \prec \pi(-2) \prec \pi(-1) \prec 
\pi(0) \succ \pi(1) \succ \pi(2) \succ \cdots,
\label{interlace relations}
\end{eqnarray}
where 
$\mu \succ \nu$ means the interlace relation;  
$\mu \succ \nu$ 
$\Longleftrightarrow$ 
$\mu_1 \geq \nu_1 \geq \mu_2 \geq \nu_2 
\geq \mu_3 \geq \cdots$.

The hamiltonian picture emerges from the above relations, 
by viewing a plane partition as evolutions of partitions 
by the discrete time $m$. 
Eventually it is described \cite{Ok-Res}
by using $2d$ free complex fermions $\psi,\psi^*$. 
We may separate the relations (\ref{interlace relations}) 
into two parts, each describing  
the evolutions for $m \leq 0$ and $m \geq 0$. 
These two types of the evolutions are 
realized in the $2d$ CFT 
by using operators $G_{\pm}$ of the forms \cite{Ok-Res}
\begin{eqnarray}
G_{\pm}=
e^{
\sum_{k=1}^{\infty}
\frac{q^{\frac{k}{2}}}{k(1-q^k)}
J_{\pm k}}, 
\label{G_pm}
\end{eqnarray}   
where $J_{\pm k}=\sum_{n=-\infty}^{\infty}:\psi_{\pm k-n}\psi^*_n:$ 
are the modes of the $U(1)$ current.

Using the free fermion description, 
one can express the generating function as 
\begin{eqnarray}
\mathcal{Z}^{U(1)}_{-i\hbar,i\hbar}(t)
=
\langle 0 |
G_+ 
Q^{L_0} e^{\frac{1}{\hbar^2}\sum_{k}t_k \hat{\mathcal{O}}_k} 
G_- 
|0\rangle\,, 
\label{fermionic representation}
\end{eqnarray}
where $L_0=\sum_{n=-\infty}^{\infty}n:\psi_{-n}\psi_n^*:$ is 
an element of the Virasoro algebra.   
The loop operators $\mathcal{O}_k$ 
are converted to operators $\hat{O}_k$ in the above representation. 
They are fermion bilinears given by 
\begin{eqnarray}
\hat{\mathcal{O}}_k=
(1-q^{-k})\sum_{n=-\infty}^{+\infty}
q^{kn}:\psi_{-n}\psi^*_{n}: 
+1\,, 
\label{hat O_k}
\end{eqnarray}

\subsection{\normalsize 
The integrable structure}

The fermion bilinears $\hat{\mathcal{O}}_k$ can be regarded as  
a commutative sub-algebra of 
the quantum torus Lie algebra 
realized by the free fermions \cite{Nakatsu-Takasaki}.
The adjoint actions of $G_{\pm}$ 
on the Lie algebra generate automorphisms of the algebra. 
Among them, 
taking advantage of 
the shift symmetry, 
the representation (\ref{fermionic representation}) 
can be eventually reformulated \cite{Nakatsu-Takasaki} to 
\begin{eqnarray}
\mathcal{Z}^{U(1)}_{-i\hbar,i\hbar}(t)
\,=\, 
\langle 0 |\,
e^{\frac{1}{2\hbar^2}\sum_{k=1}^{\infty}(-)^k(1-q^{-k})t_kJ_k}\,\, 
{\bf g}_{\star}^{5d\,U(1)}\,\, 
e^{\frac{1}{2\hbar^2}\sum_{k=1}^{\infty}(-)^k(1-q^{^k})t_kJ_{-k}}\,
| 0 \rangle\,. 
\label{toda tau}
\end{eqnarray}
In the above formula, 
${\bf g}_{\star}^{5d\,U(1)}$ is an element of $GL(\infty)$ 
of the form
\begin{eqnarray}
{\bf g}_{\star}^{5d\,U(1)}=
q^{\frac{W}{2}}G_-G_+Q^{L_0}G_-G_+q^{\frac{W}{2}}\,, 
\label{g_U(1)}
\end{eqnarray}
where $W=W_0^{(3)}=\sum_{n=-\infty}^{\infty}n^2:\psi_{-n}\psi_n:$ 
is a special element of $W_{\infty}$ algebra. 
The loop operators $\mathcal{O}_k$ are converted 
to $J_k$ or $J_{-k}$ in (\ref{toda tau}). 
These two are actually equivalent in the formula,  
since 
${\bf g}_{\star}^{5d\,U(1)}$ 
satisfies \cite{Nakatsu-Takasaki}
\begin{eqnarray}  
J_k\,{\bf g}_{\star}^{5d\,U(1)}={\bf g}_{\star}^{5d\,U(1)}J_{-k}\,, 
\hspace{6mm} 
\mbox{for}~ k \geq 0.
\label{reduction to 1-toda}
\end{eqnarray}

Viewing the coupling constants $t$ 
as a series of time variables, 
the right hand side of (\ref{toda tau}) 
is the standard form 
of a tau function of $2$-Toda hierarchy 
\cite{Ueno-Takasaki}. 
However, 
by virtue of (\ref{reduction to 1-toda}), 
the two-sided time evolutions of $2$-Toda hierarchy 
degenerate to one-sided time evolutions. 
This precisely gives the reduction to $1$-Toda hierarchy. 
Thus 
the generating function becomes a tau function of 
$1$-Toda hierarchy.

\section{\large 
Extended Seiberg-Witten geometry of $5d$ theory}

We consider the field theory limit of the $U(1)$ theory, 
which is achieved by letting $\hbar \rightarrow 0$ 
and amounts to the thermodynamic limit of the melting crystal model.
The system is described by the prepotential 
$\mathcal{F}^{(0)}(t;\Lambda,R)$.  
From the integrable system viewpoint, 
$\mathcal{F}^{(0)}$ may be interpreted as a dispersion-less tau function, 
since the generating function is substantially a tau function of $1$-Toda
hierarchy and $\mathcal{F}^{(0)}$ gives 
the leading order part of the $\hbar$ expansion of 
$\log \mathcal{Z}^{U(1)}_{-i\hbar,i\hbar}(t)$
To obtain the semi-classical solution, 
one actually needs to solve the related variational problem, 
which is reformulated as a Riemann-Hilbert problem.
This issue is treated in \cite{Nakatsu-Noma-Takasaki}.

\subsection{\normalsize 
Seiberg-Witten curve of $U(1)$ theory}

Let us present the Seiberg-Witten curve for the $U(1)$ theory. 
We first employ the following curve \cite{Maeda-Nakatsu, MNTT2}:
\begin{eqnarray}
\mathcal{C}_{\beta} :\hspace{8mm}
y+y^{-1}=
\frac{1}{R\Lambda}
(e^{-Rz}-\beta)\,,
\hspace{5mm}
z \in \mathbb{C}\,,
\label{C_beta}
\end{eqnarray}
where $\beta$ is a real parameter. 
$\mathcal{C}_{\beta}$ is a double cover of the cylinder 
$\mathbb{C}^*=\mathbb{C}/\frac{2\pi i}{R}$, 
with a cut $I$ along the real axis on the Riemann sheet. 
The coupling constants $t$ determine $\beta$ as 
$\beta=\beta(t)$. 
To see this, 
let us introduce a meromorphic 
differential of the form
\begin{eqnarray}
d\Psi=
\Big\{
1-\frac{1}{2}R^2\sum_{k=1}^{\infty}k^3t_kM_k(z)
\Big\}d\log y\,,
\end{eqnarray}
where 
$M_k(z)=\sum_{n=0}^{k}d_{k-n}(\beta)e^{-nRz}$. 
The coefficients $d_n(\beta)$ are given in  
the asymptotic expansion  
\begin{eqnarray}
\sqrt{(1-\beta e^{Rz})^2-(2R\Lambda e^{Rz})^2}
=\sum _{n=0}^{\infty}d_n(\beta)e^{nRz}\,,
\hspace{6mm}
\Re z \rightarrow -\infty.
\end{eqnarray}
Finally, 
solving the Riemann-Hilbert problem, 
$\beta$ is determined 
by the condition \cite{Nakatsu-Noma-Takasaki}
\begin{eqnarray}
\oint_{C}z d\Psi=0
\hspace{8mm}
(\mbox{$C$: a contour encircling $I$ anticlockwise}).
\label{RH condition}
\end{eqnarray}

\subsection{\normalsize 
Vevs of the loop operators}

The vev of the loop operators $\mathcal{O}_k$ can be represented 
by using an analogue of the Seiberg-Witten differential. 
Eventually, the vev can be organized to the contour integral 
\begin{eqnarray}
\frac{\partial \mathcal{F}^{(0)}(t;\Lambda,R)}{\partial t_k}
=
\lim_{\hbar \rightarrow 0}
\langle \mathcal{O}_k \rangle 
=
\frac{-kR}{2\pi i} 
\oint_C 
e^{-kRz}dS,, 
\label{vev O_k}
\end{eqnarray}
where $dS=S'(z)dz$ is an analogue of the Seiberg-Witten differential. 
$S'(z)$ is given by the indefinite integral 
\begin{eqnarray}
S'(z)=
\int^z d\Psi\,.
\label{SW differential}
\end{eqnarray}
The contour integral in the right hand side of (\ref{vev O_k}) 
can be converted to a residue integral. 
Actually, 
by using coordinate $Z=e^{-Rz}$, we obtain
\begin{eqnarray}
\frac{\partial \mathcal{F}^{(0)}(t;\Lambda,R)}{\partial t_k}
=
\lim_{\hbar \rightarrow 0}
\langle \mathcal{O}_k \rangle 
=
kR\, \mbox{Res}_{Z=\infty}
\Big(
Z^k dS
\Big)\,.
\label{vev O_k residue}
\end{eqnarray}

\subsubsection*{\underline{Acknowledgements}}
This article is based on a talk presented 
at the international workshop  
{\it ``Progress of String Theory and Quantum Field Theory''} 
(Osaka City University, December 7-10, 2007). 
We would like to thank the organizers of the conference for 
arranging such a wonderful conference.
K.T is supported in part by Grant-in-Aid for Scientific Research 
No. 18340061 and No. 19540179.


\end{document}